# Equivalent circuit theory of radiative heat transfer on micro- and nanoscale

Stanislav I. Maslovski[1,2], Constantin R. Simovski[3], Sergei A. Tretyakov[3]

[1]Instituto de Telecomunicações, Electrical Engineering Dept., Coimbra Univ., Coimbra, Portugal
[2]National Research Univ. ITMO, St. Petersburg, Russia
[3]Radio Science and Engineering Dept., Aalto Univ., Espoo, Finland
stanislav.maslovski@gmail.com

**ABSTRACT**

Here, we outline a theory of radiative heat transfer based on an equivalent electrical network representation for the hot material slabs in an arbitrary multilayered environment with arbitrary distribution of temperatures and electromagnetic properties among the layers. Our approach is fully equivalent to the known theories operating with the fluctuating current density, while being significantly simpler in analysis and applications.

## 1. Introduction

As is known, it is the fluctuation-dissipation theorem (FDT) that constitutes the basis of the present-day radiative heat transfer theories. In essence, this theorem demands that, at a finite temperature, in any material with loss there exist thermal fluctuating currents which are responsible for the thermal radiation. Such a picture places the radiative heat transfer calculations in frames of the classical electromagnetic theory rooted in the macroscopic Maxwell equations. Historically, however, thermal agitation of fluctuating currents was first discovered in electric circuits. The theory of such thermal fluctuations was developed by Nyquist [1] using only pure circuit- and transmission-line theory methods.

The result of Nyquist states that, in any linear passive two-pole network operating at a temperature $T$ the electric thermal fluctuations (i.e., the thermal noise) concentrated within an angular frequency interval $\Delta\omega$ can be equivalently represented by the fluctuating electromotive force (EMF) $e(t)$, with the mean-square of fluctuations

$$\langle e^2 \rangle = 4\Theta(\omega, T) R(\omega) \frac{\Delta\omega}{2\pi}, \qquad (1)$$

where $\Theta(\omega, T) = \hbar\omega / (\exp[\hbar\omega/k_\text{B}T] - 1)$ is Planck's energy of a harmonic oscillator, and $R(\omega)$ is the input resistance of the two-pole. The equivalent EMF is (virtually) connected in series with the two-pole, which is then considered noiseless.

The beauty of this classic result is in that *no knowledge of the internal structure* of the electric network is required, and that all the information is contained within just a single parameter (the frequency-dependent input resistance). Later, when the FDT in its electromagnetic formulation through the fluctuating current density appeared as a generalization of the Nyquist theory to the volumetric media with loss, this important and simple physical fact became somehow hidden within what seemed a more general, but significantly more cumbersome theory.

Partly, it could happen also because the concept of input impedance and the equivalent circuit description for full-wave electromagnetic problems were largely unknown among theorists working in the field of radiative heat transfer. However, in applied electromagnetics it is well-known that stratified media can be very efficiently treated within the so-called *vector transmission line theory* [2], which, in essence, assigns an equivalent transmission line network to every electromagnetic mode (propagating or evanescent) in the system. We would like to stress here that such a theory is not an approximation: It is a direct consequence of the Maxwell equations when modal expansion is applied to the electromagnetic field in layered structures.

In this work we extend the vector transmission line theory in order to include the effect of the fluctuating current density within the layers. The generalized theory allows us to prove a complete equivalence between a volumetric multilayered structure and its circuit theory counterpart, which may be visualized as a chain of transmission line segments with equivalent fluctuating voltage sources connected at the ports. We show that this equivalent network may be reduced to just a series connection of a number of voltage sources representing the fluctuating EMFs and equivalent impedances (each can be under different temperature), thus, recovering in this way the famous Nyquist result, generalized here to the full-wave electromagnetic processes in stratified media. Therefore, the calculation of the radiative heat transfer between the layers reduces in our theory to a number of equivalent circuit theory calculations, which are relatively simple and very similar to what is typically done when considering thermal noise in real electric networks.

## 2. Theory

We consider a multilayered structure composed of reciprocal material slabs and vacuum gaps. The structure is assumed translationary invariant in the plane of slabs, which allows us to expand the electromagnetic field in plane waves (both propagating and evanescent) characterized with varying transverse wave vector $\mathbf{k}_\text{t} = (k_x, k_y)$ which is the same in all layers. Throughout this text we use root-mean-square (RMS) complex amplitudes for the time-harmonic fields with the time dependence of the form $\exp(j\omega t)$.

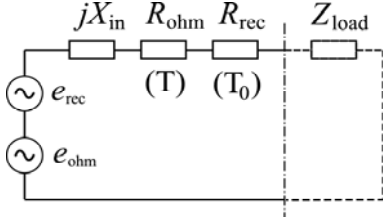

Fig. 1 Equivalent circuit of the radiative heat transfer from a material slab under a temperature $T$ backed with a background material under a temperature $T_0$.

In this work we prove that, in a given layer of the considered multilayered structure ($l$-th layer) being under the temperature $T^l$ the transverse components of the time-harmonic fluctuating electric and magnetic fields at the layer interfaces (labeled here with subscripts 1 and 2) are related as follows

$$\begin{pmatrix} \bar{\bar{Z}}^l_{11} & \bar{\bar{Z}}^l_{12} \\ \bar{\bar{Z}}^l_{21} & \bar{\bar{Z}}^l_{22} \end{pmatrix} \cdot \begin{pmatrix} \mathbf{n}_1 \times \mathbf{H}^l_{t,1} \\ \mathbf{n}_2 \times \mathbf{H}^l_{t,2} \end{pmatrix} - \begin{pmatrix} \mathbf{E}^l_{t,1} \\ \mathbf{E}^l_{t,2} \end{pmatrix} = \begin{pmatrix} \mathbf{e}^l_1 \\ \mathbf{e}^l_2 \end{pmatrix}, \quad (2)$$

where $\bar{\bar{Z}}^l_{mn} = \bar{\bar{Z}}^l_{mn}(\omega, \mathbf{k}_t)$ are the dyadic Z-parameters of the chosen layer, and $\mathbf{e}^l_{1,2}$ are the vectorial fluctuating EMFs equivalently representing the thermal fluctuations within the same layer, and $\mathbf{n}_{1,2}$ are the external unit normals at the interfaces. Eq. (2) generalizes the known result of the vector circuit theory to the case of non-vanishing thermal fluctuations. The fluctuating EMFs are related to the fluctuating thermal current density within the layer and, thus, can be found form the FDT. For the correlation of these fluctuating vectorial EMFs acting within a frequency interval $\Delta\omega$ and the transverse wave number interval $\Delta k_x \Delta k_y$ we obtain (the angular brackets denote statistical averaging)

$$\langle \mathbf{e}^l_m \cdot \mathbf{e}^{l*}_n \rangle = 4 \operatorname{Re}\left( \operatorname{Tr}\left[ \bar{\bar{Z}}^l_{mn} \right] \right) \Theta(\omega, T^l) \frac{\Delta\omega \Delta k_x \Delta k_y}{(2\pi)^3}, \quad (3)$$

which is a generalization of the Nyquist formula to the vectorial case. Here, Tr[...] denotes the trace of a dyadic. For uniaxial magnetodielectric layers when the fields split into independent polarizations (TE and TM), one obtains a similar result expressed through scalar Z-parameters of the layer for a specific polarization.

Unfortunately, due to limited space of an abstract it is impossible to give here the details on application of the vector transmission line model outlined above to the case of a general multilayered structure with arbitrary distribution of the temperatures among the layers. Below we consider a simplified example, in which the total structure can be split in two parts, where one part can be represented by a hot layer above an infinite and uniform background and the other part can be seen as an effective load for the power of thermal radiation produced in the first part.

For such a case we prove that the effect of thermal fluctuations of a given polarization in a material slab under a temperature $T$ backed with an (arbitrary) infinite background medium under a temperature $T_0$ may be represented with the equivalent circuit shown in Fig. 1. In this equivalent circuit, the backed layer is described with its total complex input impedance

$$Z_{in} = R_{in} + jX_{in} = Z_{11} - \frac{Z^2_{12}}{Z_{22} + Z_{back}}, \quad (4)$$

where $Z_{back}$ is the input impedance of the background. In Fig. 1, $R_{ohm} = R_{in} - R_{rec}$ corresponds to the part of the input resistance that is due to the ohmic loss in the layer. This part of the resistance is effectively under the temperature $T$ of the layer, and is associated with the fluctuating EMF $e_{ohm}$ such that

$$\langle |e_{ohm}|^2 \rangle = 4 R_{ohm} \Theta(\omega, T) \frac{\Delta\omega \Delta k_x \Delta k_y}{(2\pi)^3}. \quad (5)$$

The other part of the input resistance of the layer, $R_{rec}$, represents the effect of loss in the background material. It can be expressed as

$$R_{rec} = \frac{|Z_{12}|^2 \operatorname{Re}(Z_{back})}{|Z_{22} + Z_{back}|^2}. \quad (6)$$

We prove that this resistance is effectively under the temperature of the background, $T_0$, and, thus, is related to the thermal radiation received from the background. The fluctuating EMF $e_{rec}$ associated with it is such that

$$\langle |e_{rec}|^2 \rangle = 4 R_{rec} \Theta(\omega, T_0) \frac{\Delta\omega \Delta k_x \Delta k_y}{(2\pi)^3}. \quad (7)$$

The two EMFs are uncorrelated: $\langle e_{ohm} e^*_{rec} \rangle = 0$.

One may notice that very similar concepts exist also in antenna theory where the thermal noise of an antenna is represented as a sum of the noise generated locally by the ohmic losses in the antenna and the noise received from the environment. The first addend in this case is proportional to the antenna loss resistance (which vanishes for an antenna made of a perfect conductor) and the second term is proportional to the radiation resistance of the antenna.

This simple equivalent circuit allows us to calculate the radiative heat flow *from* such a backed slab *into* the adjacent half-space of an arbitrary structure (the radiative heat flow in the opposite direction can be calculated reversing the roles of the two adjacent half-spaces). Namely, within the frames of our theory, this half-space plays a role of a load of certain input impedance $Z_{load}$ (see Fig. 1). Then, the power density delivered to this half-space by the thermal fluctuations concentrated around the frequency $\omega$ and the wave vector $\mathbf{k}_t$ verifies

$$\Pi_{\omega, \mathbf{k}_t} = \frac{\langle |e_{ohm}|^2 \rangle + \langle |e_{rec}|^2 \rangle}{|Z_{in} + Z_{load}|^2} \operatorname{Re}(Z_{load}). \quad (8)$$

The total heat power transfer at a given frequency is found by summing up the contributions of all plane waves characterized with different transverse wave numbers and polarizations.

**3. Numerical examples**

The theory outlined above can be applied to arbitrary layered structures, including the ones in which some of the layers or all the layers are formed by nanostructured metamaterials. It has been recently demonstrated [3] that with the use of carbon nanotubes or metallic nanowires it is possible to achieve a giant radiative heat transfer over rather thick gaps, due to the fact that such metamaterials can transform the evanescent waves of free space into the propagating modes of the metamaterial. At the same time, by nanostructuring the metamaterial it is possible to keep the amount of heat transferred by phonons at a low level.

Here, we calculate the radiative heat transfer in a structure composed of three adjacent layers. The outer layers (1 and 3) are assumed to be infinitely thick. The middle layer 2 is a vacuum gap of a finite thickness. The heat is transferred from the hot layer 1 to the cold layer 3. We consider a few characteristic cases in which metallic nanorods are embedded into (some of) the layers. The materials of the layers are either Copper Indium Gallium Sulfur-selenide (CIGSS) or germanium (Ge). The numerical results are represented in Figs. 2, 3.

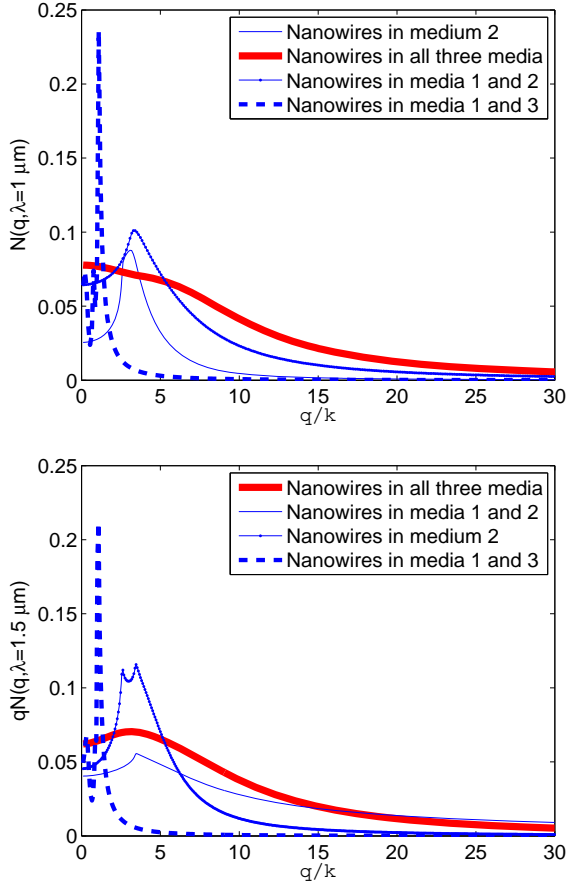

Fig. 2 Spatial spectrum of the heat transfer function between medium 1 (SiC) and medium 3 across the gap $d = 2$ μm versus dimensionless spatial frequency ($q/k$). (top) Medium 3 is CIGSS. (bottom) Medium 3 is Ge.

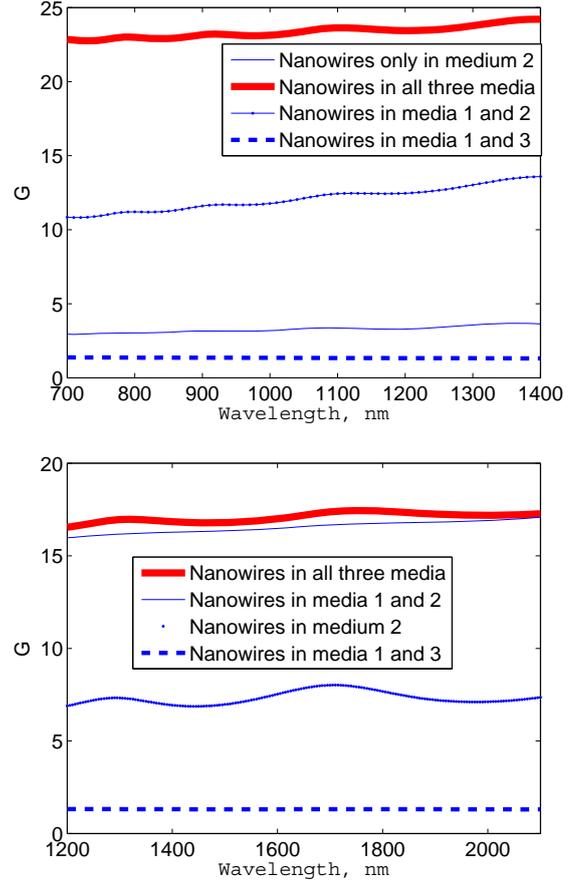

Fig. 3 Gain in the heat transfer power versus wavelength λ for the design solutions with nanowires compared to that in the case when nanowires are absent. (top) Medium 3 is CIGSS. (bottom) Medium 3 is Ge.

The layers with embedded nanorods where modeled as indefinite dielectric media. In these examples we optimize the volume fraction of the nanorods (made of tungsten in these examples) in order to achieve higher amount of power transferred by the radiation. The best design solution corresponds to nanowires located in all three media. More details regarding these results will be given in the presentation.

**4. Concluding remarks**

In the presentation we discuss further generalizations of the theory outlined above, as well as its practical application to the radiative heat transfer on micro- and nanoscale in photovoltaic devices and systems. We show that the equivalent circuit theory allows us to use the well-known circuit theory principles, for example, when studying the problem of maximizing the heat transfer by means of electromagnetic radiation.